\documentclass[aps, prx, twocolumn, superscriptaddress, ]{revtex4-1}
\usepackage{amsmath,amssymb,amsfonts,amsbsy}
\usepackage{graphicx}
\usepackage{subfig}
\graphicspath{{./figures/}}
\usepackage{dcolumn}
\usepackage{bm}
\usepackage{multirow}
\usepackage{mathtools}
\usepackage{array}
\usepackage{color}
\usepackage[normalem]{ulem}
\usepackage[per-mode=symbol]{siunitx}
\usepackage{upgreek}

\usepackage[labelfont=bf]{caption}
\usepackage[figurename=Fig.]{caption}

\DeclareSIUnit \belm {Bm}

\usepackage[hidelinks]{hyperref} % should be before cleveref
\usepackage[capitalise,nameinlink]{cleveref} % should always be the last package
\usepackage{booktabs}
\usepackage[utf8]{inputenc}
\usepackage[T1]{fontenc}

\newcommand{\app}[1]{\hyperref[app:#1]{Appendix~\ref*{app:#1}}}

\newcommand{\ket}[1]{\left|#1\right\rangle}

\def\@setaltaffiliation{\vspace{-\baselineskip}\def\altaffiliation##1{\@par##1\@addpunct.}\altaffiliationes}
\def\@setaltaffiliation{\vspace{-\baselineskip}\def\altaffiliation##1{\@par##1\@addpunct.}\altaffiliationes}
\RequirePackage{lineno}
\setpagewiselinenumbers
\begin{document}
\title{CMOS-based cryogenic control of silicon quantum circuits}

\author{Xiao~Xue}
 \altaffiliation{These authors contributed equally to this work;}
\affiliation{QuTech, Delft University of Technology, Lorentzweg 1, 2628 CJ Delft, Netherlands}
\affiliation{Kavli Institute of Nanoscience, Delft University of Technology, 2600 GA Delft, Netherlands}
\author{Bishnu~Patra}
 \altaffiliation{These authors contributed equally to this work;}
\affiliation{QuTech, Delft University of Technology, Lorentzweg 1, 2628 CJ Delft, Netherlands}
\affiliation{Kavli Institute of Nanoscience, Delft University of Technology, 2600 GA Delft, Netherlands}
\affiliation{Department of Quantum and Computer Engineering, Delft University of Technology, Mekelweg 4, 2628CD Delft, Netherlands}
\author{Jeroen~P.~G.~van~Dijk}
 \altaffiliation{These authors contributed equally to this work;}
\affiliation{QuTech, Delft University of Technology, Lorentzweg 1, 2628 CJ Delft, Netherlands}
\affiliation{Kavli Institute of Nanoscience, Delft University of Technology, 2600 GA Delft, Netherlands}
\affiliation{Department of Quantum and Computer Engineering, Delft University of Technology, Mekelweg 4, 2628CD Delft, Netherlands}
\author{Nodar~Samkharadze}
\affiliation{QuTech, Delft University of Technology, Lorentzweg 1, 2628 CJ Delft, Netherlands}
\affiliation{Netherlands Organization for Applied Scientific Research (TNO), Stieltjesweg 1, 2628 CK Delft, Netherlands}
\author{Sushil~Subramanian}
\affiliation{Intel Corporation, Hillsboro, OR 97124, USA}
\author{Andrea~Corna}
\affiliation{QuTech, Delft University of Technology, Lorentzweg 1, 2628 CJ Delft, Netherlands}
\affiliation{Kavli Institute of Nanoscience, Delft University of Technology, 2600 GA Delft, Netherlands}
\author{Charles~Jeon}
\author{Farhana~Sheikh}
\affiliation{Intel Corporation, Hillsboro, OR 97124, USA}

\author{Esdras~Juarez-Hernandez}
\author{Brando~Perez~Esparza}
\affiliation{Intel Guadalajara, 45017 Zapopan, Mexico}
\author{Huzaifa~Rampurawala}
\author{Brent~Carlton}
\author{Surej~Ravikumar}
\author{Carlos~Nieva}
\author{Sungwon~Kim}
\author{Hyung-Jin~Lee}
\affiliation{Intel Corporation, Hillsboro, OR 97124, USA}

\author{Amir~Sammak}
\affiliation{QuTech, Delft University of Technology, Lorentzweg 1, 2628 CJ Delft, Netherlands}
\affiliation{Netherlands Organization for Applied Scientific Research (TNO), Stieltjesweg 1, 2628 CK Delft, Netherlands}
\author{Giordano~Scappucci}
\author{Menno~Veldhorst}
\affiliation{QuTech, Delft University of Technology, Lorentzweg 1, 2628 CJ Delft, Netherlands}
\affiliation{Kavli Institute of Nanoscience, Delft University of Technology, 2600 GA Delft, Netherlands}

\author{Fabio~Sebastiano}
 \altaffiliation{These authors contributed equally to this work}
 \altaffiliation{\\Correspondence authors: \\edoardo.charbon@epfl.ch; l.m.k.vandersypen@tudelft.nl}
\affiliation{QuTech, Delft University of Technology, Lorentzweg 1, 2628 CJ Delft, Netherlands}
\affiliation{Department of Quantum and Computer Engineering, Delft University of Technology, Mekelweg 4, 2628CD Delft, Netherlands}

\author{Masoud~Babaie}
 \altaffiliation{These authors contributed equally to this work}
 \altaffiliation{\\Correspondence authors: \\edoardo.charbon@epfl.ch; l.m.k.vandersypen@tudelft.nl}
\affiliation{QuTech, Delft University of Technology, Lorentzweg 1, 2628 CJ Delft, Netherlands}
\affiliation{Department of Quantum and Computer Engineering, Delft University of Technology, Mekelweg 4, 2628CD Delft, Netherlands}

\author{Stefano~Pellerano}
 \altaffiliation{These authors contributed equally to this work}
 \altaffiliation{\\Correspondence authors: \\edoardo.charbon@epfl.ch; l.m.k.vandersypen@tudelft.nl}
\affiliation{Intel Corporation, Hillsboro, OR 97124, USA}

\author{Edoardo~Charbon}
 \altaffiliation{These authors contributed equally to this work}
 \altaffiliation{\\Correspondence authors: \\edoardo.charbon@epfl.ch; l.m.k.vandersypen@tudelft.nl}

\affiliation{École Polytechnique Fédérale de Lausanne (EPFL), 1015 Lausanne, Switzerland}
\affiliation{QuTech, Delft University of Technology, Lorentzweg 1, 2628 CJ Delft, Netherlands}
\affiliation{Kavli Institute of Nanoscience, Delft University of Technology, 2600 GA Delft, Netherlands}
\affiliation{Intel Corporation, Hillsboro, OR 97124, USA}

\author{Lieven~M.~K.~Vandersypen}
 \altaffiliation{These authors contributed equally to this work}
 \altaffiliation{\\Correspondence authors: \\edoardo.charbon@epfl.ch; l.m.k.vandersypen@tudelft.nl}

\affiliation{QuTech, Delft University of Technology, Lorentzweg 1, 2628 CJ Delft, Netherlands}
\affiliation{Kavli Institute of Nanoscience, Delft University of Technology, 2600 GA Delft, Netherlands}
\affiliation{Intel Corporation, Hillsboro, OR 97124, USA}
\date{\today}

\maketitle

% \section{Abstract}
% \label{sec:abstract}

\textbf{
The most promising quantum algorithms require quantum processors hosting millions of quantum bits when targeting practical applications~\cite{van2013blueprint}. A major challenge towards large-scale quantum computation is the interconnect complexity. In current solid-state qubit implementations, a major bottleneck appears between the quantum chip in a dilution refrigerator and the room temperature electronics. Advanced lithography supports the fabrication of both CMOS control electronics and qubits in silicon. When the electronics are designed to operate at cryogenic temperatures, it can ultimately be integrated with the qubits on the same die or package, overcoming the wiring bottleneck~\cite{vandersypen2017interfacing,patra2018cryo,pauka2019cryogenic,geck2019control}. Here we report a cryogenic CMOS control chip operating at 3\,K, which outputs tailored microwave bursts to drive silicon quantum bits cooled to 20\,mK. We first benchmark the control chip and find electrical performance consistent with 99.99\% fidelity qubit operations, assuming ideal qubits. Next, we use it to coherently control actual silicon spin qubits~\cite{watson2018programmable,zajac2018resonantly,huang2019fidelity} and find that the cryogenic control chip achieves the same fidelity as commercial instruments. Furthermore, we highlight the extensive capabilities of the control chip by programming a number of benchmarking protocols as well as the Deutsch-Josza algorithm~\cite{deutsch1992rapid} on a two-qubit quantum processor. These results open up the path towards a fully integrated, scalable silicon-based quantum computer.
}
% \section*{Introduction}
% \label{sec:introduction}

\begin{figure*}[htbp]
\centering
  \includegraphics[width=1\linewidth]{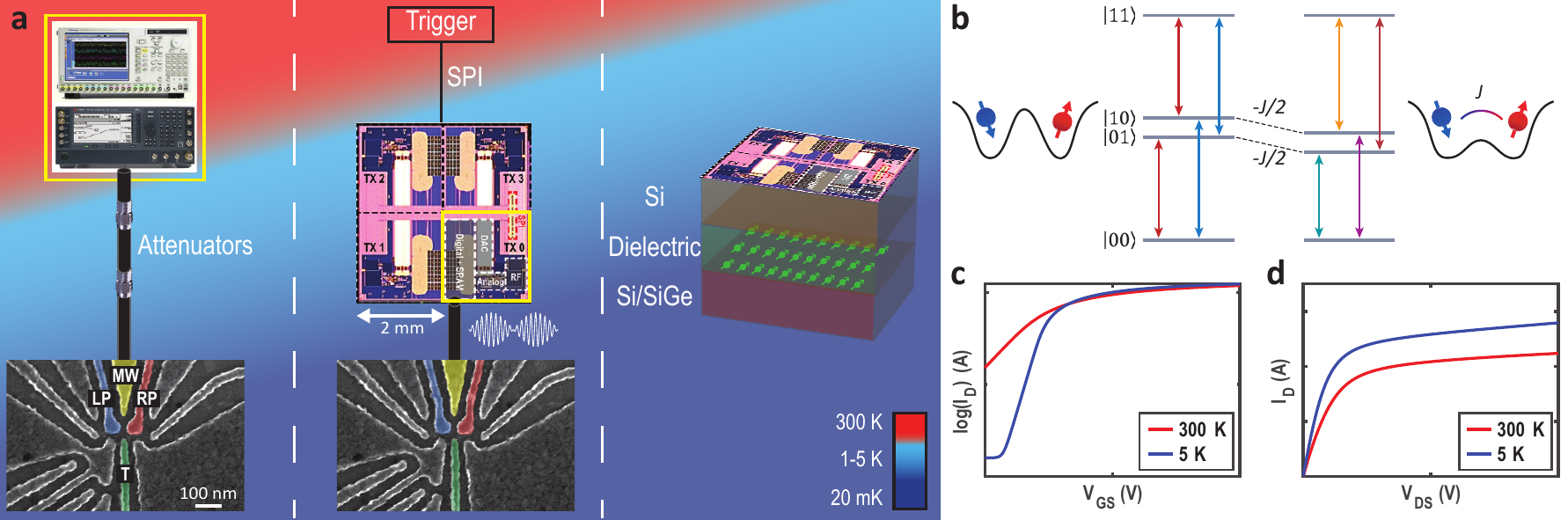}
\caption{\textbf{The cryogenic quantum control system.} \textbf{a.} Three stages of development of the control system towards full integration. From left to right: room temperature instruments connected to qubits via coax lines and attenuators; cryo-controller placed at 1-\SI{5}{\kelvin} directly connected to the qubits and triggered from room temperature using a serial peripheral interface (SPI); a future perspective of fully integrated control electronics and qubits on the same package/die. Two single electron spins used as qubits are located underneath gates LP (blue) and RP  (red), as shown in the SEM image. Multiplexed microwave signals are sent to gate MW (yellow) to control both qubits. Gate T (green) is used to tune the coupling between the qubits. \textbf{b.} Energy level diagram without (left) and with (right) exchange coupling ($J$). The resonance frequency of each qubit depends on the other qubit only when the coupling is on (low tunnel barrier between the dots). \textbf{c, d.} FinFET NMOS device characteristics at room temperature versus \SI{5}{\kelvin}: drain current (I$_D$) versus  gate-source voltage (V$_{GS}$) at drain-source voltage V$_{DS}$ = \SI{1}{\volt} (\textbf{c}) and I$_D$ versus V$_{DS}$ at V$_{GS}$ = \SI{0.4}{\volt}(\textbf{d}).}
\label{fig:setup}
\end{figure*}

A practical quantum computer comprises two main building blocks -- a quantum processor with millions of qubits and classical instrumentation to generate control signals (input) and to process readout signals (output)~\cite{vandersypen2017interfacing,van2013blueprint}. A standard setup for semiconducting or superconducting qubits has the qubits operating in a dilution refrigerator at $\sim$\SI{20}{\milli\kelvin}, while bulky microwave vector sources and arbitrary waveform generators are placed at room temperature and connected to the qubits via long cables and attenuators (\cref{fig:setup}a, left). This approach has recently enabled an experimental demonstration of the advantage of quantum computing over classical computing in a random circuit sampling experiment, that utilizes a superconducting quantum processor consisting of 53 qubits~\cite{arute2019quantum}. This system requires more than 200 coaxial control lines from room temperature to the quantum chip operated below \SI{20}{\milli\kelvin}. This brute-force approach to reach higher qubit numbers will soon hit its limits. A promising path forward is to bring the control electronics close to the quantum chip, at cryogenic temperatures~\cite{bardin2019design,patra202019,le202019,bonen2018cryogenic,esmailiyan2020fully,ekanayake2010characterization,mukhanov2019scalable,patra2018cryo,pauka2019cryogenic,xu2020chip}. Here the challenge is that the power dissipation of the control electronics easily surpasses the typical cooling power of \SI{10}{\micro\watt} available at \SI{20}{\milli\kelvin}. Silicon spin qubits are well-positioned for co-integration with dissipative classical electronics, since they can be operated above \SI{1}{\kelvin}~\cite{petit2020universal,yang2020operation}, where the cooling power is orders of magnitude higher (\cref{fig:setup}a, right). 
 Therefore, an important next step is to design and implement a quantum control chip operating at 1-\SI{3}{\kelvin}, and to test its overall performance in driving real qubits. In order to benchmark the limits of  the controller, it is advantageous to keep the qubits at $\sim$\SI{20}{\milli\kelvin}, where the qubits are most coherent and the demands on the controller are highest (\cref{fig:setup}a, middle).

A cryogenic quantum controller for practical quantum information processing must meet multiple criteria: a form factor compatible with integration in a cryogenic refrigerator; frequency multiplexing to facilitate scalability; low power consumption within the limit of refrigerator cooling power; sufficiently high output power to enable fast operations compared to the qubit coherence times; high signal-to-noise ratio (SNR) and spurious-free-dynamic-range (SFDR) for high-fidelity control; the ability to generate complex pulse shapes and perform a universal set of quantum operations; an integrated instruction set memory for the efficient execution of complex algorithms. All these requirements can be met by commercial CMOS circuits designed to operate at a few K.\par

In this work, we utilize a  quantum control chip operating at \SI{3}{\kelvin} (cryo-controller, named Horse Ridge) and fabricated in Intel \SI{22}{\nano\meter}-FinFET low-power CMOS technology~\cite{patra202019} to coherently control two electron spin qubits in a silicon double quantum dot cooled to $\sim$\SI{20}{\milli\kelvin}. Extensive electrical characterization and benchmarking using the quantum processor show that the cryo-controller meets all the above criteria.

% \section*{Setup}
% \label{sec:setup}

\begin{figure*}[htbp] 
\center{\includegraphics[width=1\linewidth]{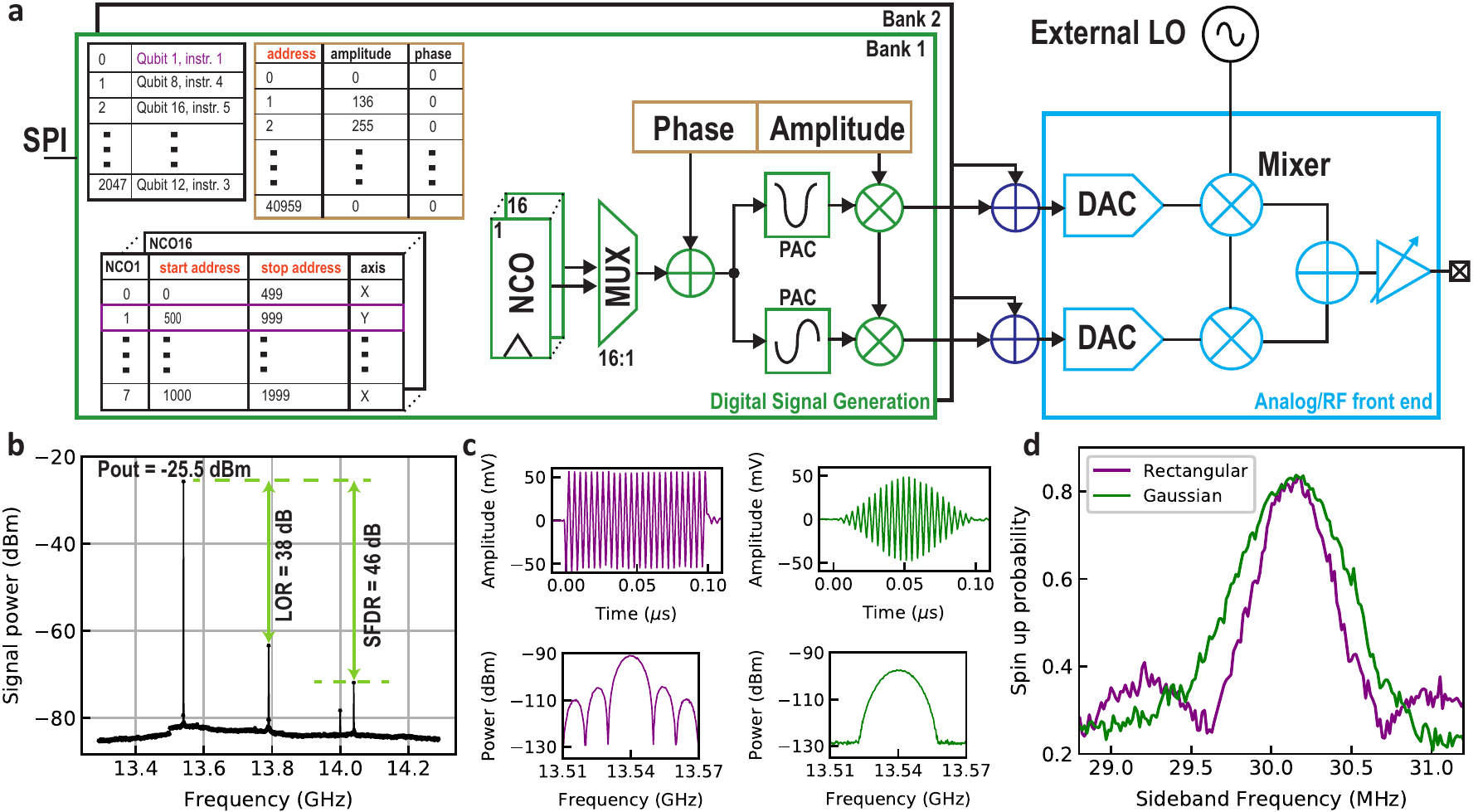}}
\caption{\textbf{The Horse Ridge cryogenic  controller characterized at \SI{3}{\kelvin}.} \textbf{a.} System-level representation showing the digital signal generation and analog/RF front end of the cryo-controller, programmable via the SPI. \textbf{b.} Continuous-wave output spectrum from the cryo-controller at \SI{13.54}{\giga\hertz} showing the main output tone, SFDR and LO rejection ratio (LOR).  \textbf{c.} Rectangular (purple) and Gaussian (green) shaped bursts before up-conversion and the corresponding spectra after up-conversion. \textbf{d.} Qubit response for different burst envelopes, obtained when sweeping the NCO frequency around the qubit resonance across a span of $\sim$\SI{3}{\mega\hertz} with a resolution of \SI{15}{\kilo\hertz}.}
\label{fig:controller}
\end{figure*}

The specifications for the cryo-controller derive from the demands on the qubit control. Here we target qubits that can be resonantly controlled with drive frequencies in the 2-\SI{20}{\giga\hertz} band, covering the typical resonance frequencies of both superconducting and spin qubits. The cryo-controller has four output ports, each with up to 32 frequency-multiplexed tones. Since the controller must dissipate minimal power and have a small form factor, we analyze in detail the signal specifications that are sufficient to achieve a $99.99\%$ gate fidelity \cite{van2020design}. Among other performance metrics, the most stringent ones dominating the architecture and power consumption of the controller are the SNR ($>$ \SI{48}{\deci\bel}) and SFDR ($>$ \SI{44}{\deci\bel}) for frequency-multiplexed control~\cite{van2020design}.

Further challenges arise in designing complex CMOS circuits at deep cryogenic temperatures. Key device characteristics such as the threshold voltage (V$_{th}$) and mobility ($\mu$) increase compared to room temperature, as seen in \cref{fig:setup}c, d \cite{beckers2018characterization}. Moreover, the degradation of active device matching \cite{hart2020mismatch} and the improvement of the quality factor of on-chip passive components \cite{patra2020characterization}, necessitate careful characterization and modeling for circuits operated at cryogenic temperatures.\par

As a benchmark of performance, we use the cryo-controller to coherently control a two-qubit quantum processor. The quantum processor is made of a double quantum dot (DQD) electrostatically confined in a $^{28}$Si/SiGe heterostructure. By tuning the voltage on plunger gates LP and RP, two single electrons are locally accumulated underneath each gate, shown in blue and red in the scanning electron microscope (SEM) image in \cref{fig:setup}a. By applying an external magnetic field of \SI{380}{\milli\tesla}, combined with the longitudinal magnetic field induced by a micro-magnet on top of the DQD (see Extended Data \cref{fig:micromagnet}), we can encode the qubit states into the Zeeman split states of the two electrons, where spin-up is used as $\ket{1}$ and spin-down is used as $\ket{0}$. The resonance frequencies of Qubit 1 ($Q_1$, underneath gate LP) and Qubit 2 ($Q_2$, underneath gate RP) are \SI{13.62}{\giga\hertz} and \SI{13.51}{\giga\hertz}, respectively. Rotations around the $\hat{x}$ and $\hat{y}$ axes are implemented by sending microwave bursts with the microwave phase controlling the rotation axis. The microwave bursts are applied to gate MW, which drives electric-dipole spin resonance (EDSR) enabled by the transverse magnetic field gradient from the micro-magnet~\cite{pioro2008electrically}, while the rotation around the $\hat{z}$ axis (phase control) is achieved by changing the reference phase in the cryo-controller~\cite{vandersypen2005nmr}. The two-qubit interaction is mediated by the exchange coupling ($J$) between the two spins~\cite{petta2005coherent}, controlled by gate T. Its effect here is to shift the anti-parallel spin states down in energy~\cite{meunier2011efficient}. As a result, the resonance frequency of each qubit now depends on the state of the other qubit, allowing conditional operations on each qubit via narrow-band microwave bursts~\cite{zajac2018resonantly,huang2019fidelity} (\cref{fig:setup}b). The corresponding four different frequencies can be individually addressed using frequency multiplexing. Both qubits are read out in single-shot mode~\cite{xue2020repetitive} (see Methods).\par

\begin{figure*}[htbp] 
\center{\includegraphics[width=1\linewidth]{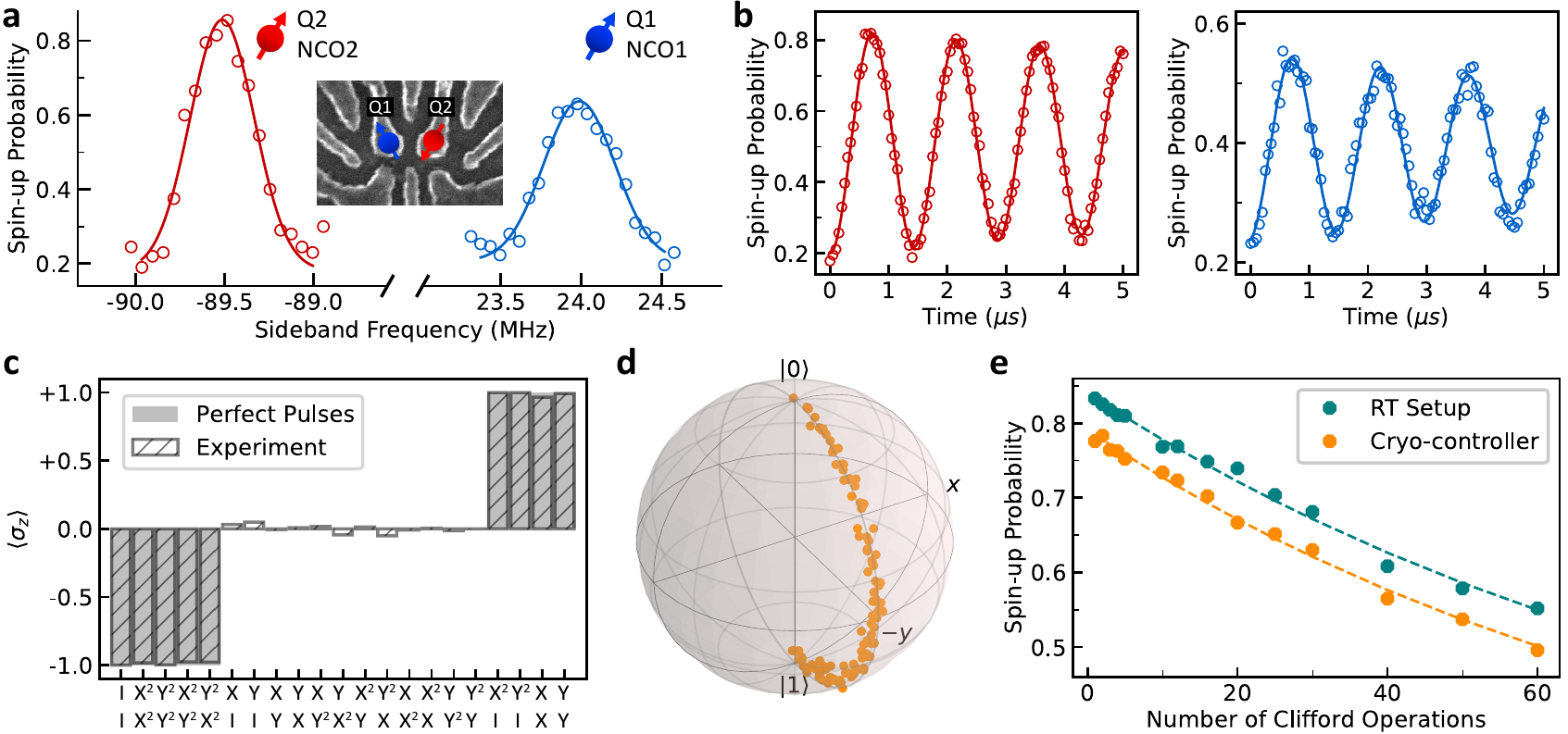}}
\caption{\textbf{Frequency-multiplexed qubit control and fidelity benchmarks with the cryo-controller} \textbf{a.} Spectra showing the qubit resonances. Inset: SEM image indicating the qubits' positions. \textbf{b.} Frequency-multiplexed control producing simultaneous Rabi oscillations of $Q_2$ (left) and $Q_1$ (right). The decay arises mainly from the residual coupling between the two qubits (see Extended Data \cref{fig:individual_rabi} for Rabi oscillations in individual driving mode). \textbf{c.} $\langle\sigma_z\rangle$ of $Q_2$ measured after an an AllXY sequence. The output power is calibrated to achieve a $\sim$\SI{1}{\mega\hertz} Rabi frequency (the same applies to the QST and RB experiments). The visibility is normalized by removing the readout error (see Methods). \textbf{d.} Trajectory of the state of $Q_2$  under an $X^2$ gate  reconstructed by QST. Orange data points indicate the qubit state after incrementing microwave burst times. \textbf{e.} Randomized benchmarking of $Q_2$ performed by the cryo-controller and the room temperature setup. We offset the orange data points by $-0.05$ to facilitate comparison of the two traces.}
\label{fig:single-qubit}
\end{figure*}

\Cref{fig:controller} shows the system-level architecture of the cryo-controller, which consists of a digital signal generation unit with an analog/RF front-end. At the core of the digital signal generation, a numerically controlled oscillator (NCO) outputs a sequence of bit strings every clock period. This bit string encodes a phase that is intended to track the reference phase of one particular qubit. The output of 16 NCOs is multiplexed and fed to a phase-to-amplitude converter (PAC) to generate a sinusoidal (in-phase) and cosinusoidal (quadrature-phase) signal. The NCO phases are constructed via a phase accumulator, which increments the phase in steps determined by a digital frequency tuning word (FTW). The 22-bit FTWs in combination with the \SI{1}{\giga\hertz} clock frequency of the phase accumulator gives a frequency resolution of $\sim$ \SI{238}{\hertz}.

The sine and cosine signals are amplitude and phase modulated using the envelope memory (orange box) containing up to 40960 points, each specifying an amplitude and phase value. An instruction table memory can store up to 8 different instructions per qubit/NCO by referring to start and stop addresses in the envelope memory. Finally, these instructions are listed in the instruction list to execute up to 2048 instructions from multiple instruction tables, initiated by a single external trigger. The output of two such banks, each generating a digital signal, are summed to simultaneously control two qubits, consequently increasing the number of supported (uncoupled) qubits from 16 to 32.\par

The generated digital signals are translated to the analog domain using high-speed digital-to-analog converters (DAC) and upconverted to the required qubit frequency using an I/Q mixer and an external local oscillator (LO). Finally, an output driver is incorporated to produce the required voltage amplitude (through a tunable gain of \SI{40}{\deci\bel}) in the frequency range of 2 to \SI{20}{\giga\hertz}, while driving the \SI{50}{\ohm} coaxial cable connecting to the qubits. Such a wide frequency and output power range allows the control of various solid-state qubits such as spin qubits and superconducting qubits. The controller dissipates \SI{384}{\milli\watt} with all the NCOs simultaneously operating at a clock frequency of \SI{1}{\giga\hertz} (Digital Signal Generation: \SI{330}{\milli\watt}, Analog/RF front-end: \SI{54} {\milli\watt}) (see Methods). This architecture is replicated 4 times in a die area of \SI{16}{\milli\meter}$^2$ (TX0-TX3 in \cref{fig:setup}) with an ability to control up to 4 $\times$ 32 frequency multiplexed qubits. \par

The purity of the generated signal can be quantified using the output signal spectrum shown in \cref{fig:controller}b. The generated signal has an SFDR of \SI{46}{\deci\bel} at \SI{13.54}{\giga\hertz} in a \SI{1}{\giga\hertz} bandwidth, excluding the residual LO leakage (see Extended Data \cref{fig:rf_output} for a two-tone test). The noise floor is flat across the \SI{1}{\giga\hertz} bandwidth, 
and the cryo-controller leaves the electron temperature of the quantum device unaffected (see Extended Data \cref{fig:electron_temperature}).
The SNR is \SI{48}{\deci\bel} when integrating over  \SI{25}{\mega\hertz}, corresponding to the targeted maximum qubit Rabi frequency. Along with the low quantization noise and frequency noise, the output signal quality is predicted to achieve a single-qubit gate fidelity of 99.99\%, assuming ideal qubits \cite{van2020design}. The amplitude and phase modulation capabilities of the controller allow the chip to generate arbitrary waveforms to precisely shape the spectral content of the pulse used to manipulate the qubits, as shown in \cref{fig:controller}c. In illustration, \cref{fig:controller}d shows the response of $Q_2$ to a microwave burst with rectangular versus Gaussian envelope, both calibrated to invert the qubit state when the drive is on-resonance with the qubit.\par

% \section*{Performance Benchmark}

\begin{figure*}[htbp] 
\center{\includegraphics[width=1\linewidth]{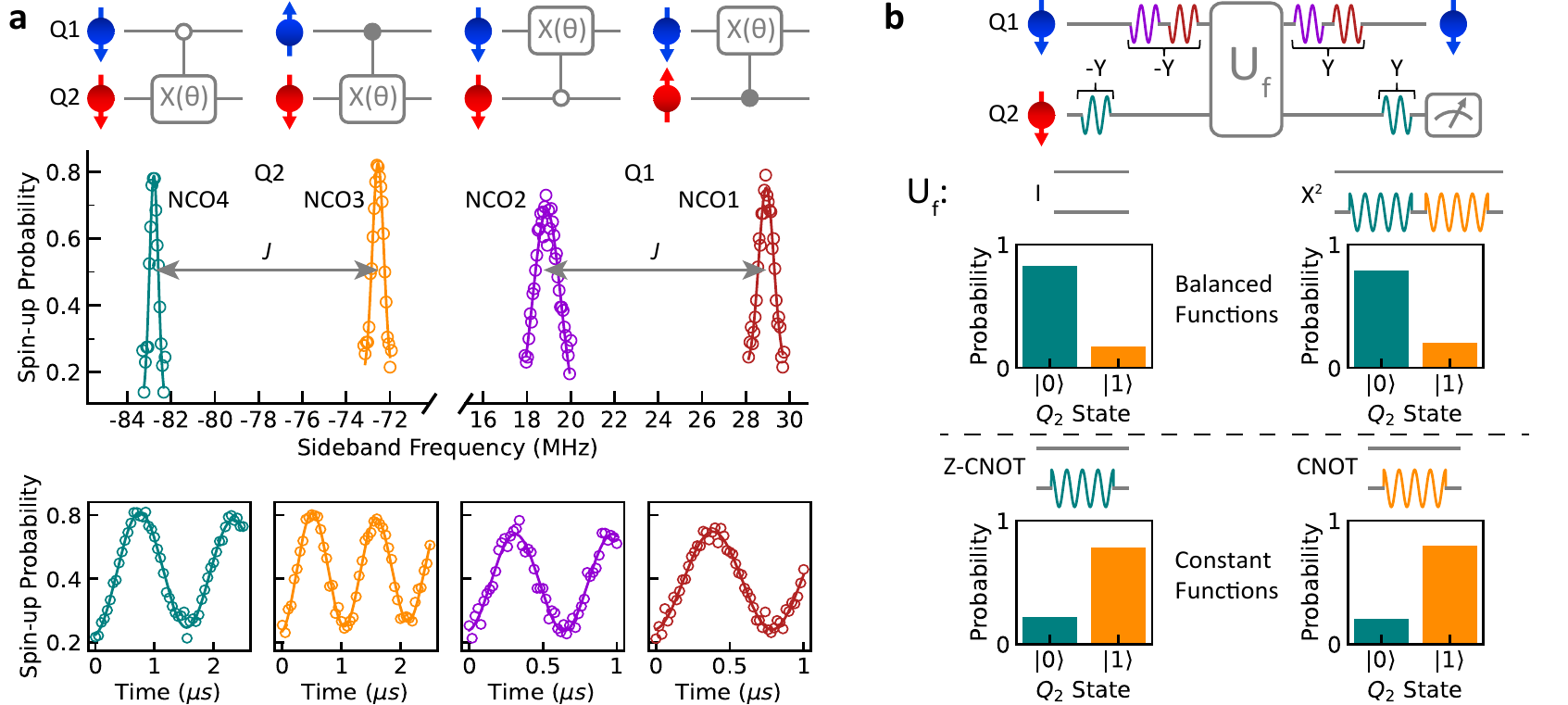}}
\caption{\textbf{Programming a quantum processor with the cryo-controller.} \textbf{a.} Two-qubit logic with the cryo-controller. The middle panel shows the  spectra of two qubits obtained using the cryo-controller with the exchange coupling ($J$) between the qubits turned on. Selective excitation of each of the four resonances can be used for implementing various two-qubit controlled-rotation gates, shown in the upper panel. The lower panels (shared Y-axis labels) show the Rabi oscillations at each frequency. \textbf{b.} (Top) Pulse sequences of the Deutsch–Josza algorithm programmed into the cryo-controller and (Bottom) measured probabilities of the output qubit state ($Q_2$) after running the algorithm. A constant function is composed of either a $CNOT$ or a $Z$-$CNOT$ gate, which consists of a $CROT$ gate on $Q_2$ and a phase correction on $Q_1$ (not plotted). Only the lower frequency (green branch, $Z$-$CROT$) is used for the $-Y$ and $Y$ gate on $Q_2$ because $Q_1$ (ideally) starts from and ends up in $\ket{0}$. The visibility of $Q_2$ is normalized by removing the readout error. Empirically, we attribute the remaining errors mostly to charge noise in the presence of a finite $J$ (see Methods).}
\label{fig:two-qubit}
\end{figure*}

Next, we test the functionality of the cryo-controller for controlling uncoupled qubits. The LO frequency is set to \SI{13.54}{\giga\hertz}. $Q_1$ is then offset from the LO by \SI{24}{\mega\hertz} and $Q_2$ by \SI{-90}{\mega\hertz}. The qubit resonances are found by sweeping one single-sideband tone generated by one NCO~(\cref{fig:single-qubit}a), using the 22-bit FTW. Then we use one NCO from each bank to generate two tones on resonance with the two qubits and drive simultaneous Rabi oscillations on both qubits~(\cref{fig:single-qubit}b). Here a \SI{5}{\micro\second} rectangular envelope is uploaded to the envelope memory, and saved as an instruction. The duration of the microwave burst is swept by updating the start or stop address of this instruction.\par

The pulses for single-qubit rotations are precisely calibrated using the AllXY sequence~\cite{reed2013entanglement}. In the AllXY experiment, 21 different pairs of single-qubit gates from the set $\{I, X, Y, X^2, Y^2\}$ are applied to a qubit initialized to $\ket{0}$. Here $I$ is the identity operation, $X$ and $Y$ are ${\pi}/2$ rotations around the $\hat{x}$ and $\hat{y}$ axis respectively, and $X^2$ and $Y^2$ are $\pi$ rotations. The final state $\hat{z}$-projection  $\langle\sigma_z\rangle$ takes values from $\{-1,0,+1\}$ under perfect operations (shown as the gray shaded areas in \cref{fig:single-qubit}c). Any miscalibration in the amplitude, frequency or phase of the pulse results in deviations from the ideal outcome (hatched bars in \cref{fig:single-qubit}c). In addition, we reconstruct the trajectory of an $X^2$ gate by performing quantum-state tomography (QST)~\cite{altepeter2005photonic} at incremental burst times of a rectangular microwave signal (\cref{fig:single-qubit}d). The AllXY and QST results indicate that the single-qubit gate set is well calibrated, offering a good starting point for benchmarking the gate fidelity.\par

The gate fidelity is a crucial metric to express the performance of a quantum processor and its classical controller. We use single-qubit randomized benchmarking (RB)~\cite{knill2008randomized, magesan2012characterizing} to compare the performance of the cryo-controller with the conventional room temperature (RT) setup, which consists of an arbitrary waveform generator (Tektronix 5014C) and a vector signal generator (Keysight E8267D). A programmable microwave switch placed at the \SI{3}{\kelvin} plate allows to conveniently alternate between the cryo-controller and the RT setup. In the RB experiment, sequences of increasing numbers of randomly selected Clifford operations are applied to the qubit ($Q_2$), followed by a final Clifford operation that returns the qubit to its initial state in the ideal case. For each data point in \cref{fig:single-qubit}e, 32 different sequences are randomly sampled and each is repeated 200 times. Envelopes of all gates to be used are uploaded to the envelope memory, and saved as instructions. The random sequences are constructed by updating the instruction list. The instructions in the list are executed sequentially after an external trigger via the SPI in \cref{fig:controller}a is received. Exactly the same random sequences are used in an RB experiment using the RT setup. We find an average single-qubit gate fidelity of $99.71\pm0.03\%$ with the RT setup and $99.69\pm0.02\%$ with the cryo-controller~(see Methods). The fidelities are consistently identical within the error bars and well above the threshold for fault-tolerance~\cite{fowler2012surface}, with the infidelity limited by the qubit. These experiments demonstrate the high signal quality from the cryo-controller as well as its capability of generating complex sequences.\par

% \section*{Cryogenic Controlled Quantum Processor}

To further test the programmability of the cryo-controller, we use it to implement two-qubit logic in the quantum processor. Taking advantage of the frequency shift of each qubit conditional on the state of the other qubit (\cref{fig:setup}b), we use controlled-rotation ($CROT$) gates as the native two-qubit gates. These are achieved by frequency selective addressing~\cite{zajac2018resonantly, huang2019fidelity}, thus demanding 2 NCOs per qubit (see Methods). A $\pi$-rotation at the higher or lower frequency implements the canonical controlled-NOT ($CNOT$) gate or the zero-controlled-NOT ($Z$-$CNOT$) gate respectively, up to a single-qubit $\pi/2$ $\hat{z}$-rotation on the control qubit. Due to cross-talk, an additional phase correction in the form of a $\hat{z}$-rotation is needed. All $\hat{z}$ rotations are implemented by updating the reference phase of the NCO (see Extended Data \cref{fig:schematic}).
Single-qubit gates are implemented by addressing both frequencies of the same qubit sequentially. Making use of four NCOs, we program the cryo-controller to run the two-qubit Deutsch–Josza algorithm, which determines whether a function gives constant or balanced outcomes~\cite{deutsch1992rapid}. The two constant (balanced) functions that map one input bit on one output bit are implemented by the $CNOT$ and $Z$-$CNOT$ ($I$ and $X^2$) operations. 
Here, we choose $Q_1$ to be the output qubit and $Q_2$ to be the input qubit. \cref{fig:two-qubit}b shows the pulse sequence and the measurement results, where the constant (balanced) functions lead to a high probability for measuring the data qubit as $\ket{1}$ ($\ket{0}$), as expected. This experiment highlights the ability to program the cryo-controller with arbitrary sequences of operations. 

The cryo-controller allows for much more complex sequences, containing up to 2048 instructions for each of the four transmitters. Each instruction defines a microwave burst at one of 32 independent frequencies with an amplitude and phase profile that can be arbitrarily shaped. The cryo-controller can be conveniently embedded in existing micro-architectures and programmed via standard QASM variants\cite{svore2006layered}. This quantum-classical architecture can thus be directly applied to multi-qubit algorithms and noisy intermediate-scale quantum devices ~\cite{preskill2018quantum}.

% \section*{Conclusion}

The versatile programmability combined with the signal quality allowing up to $99.99\%$ gate fidelities, the footprint of just \SI{4}{\square\milli\meter}, the power consumption of \SI{384}{\milli\watt}, the ability to integrate multiple transmitters on one die, and operation at \SI{3}{\kelvin}, highlight the promise of cryo-controllers to address key challenges in building a large-scale quantum computer. 
Optimized design of cryogenic CMOS circuits, e.g. the use of a narrower frequency band, can substantially reduce the power consumption~(see Methods) and make it possible to work at \SI{1}{\kelvin} or even lower temperatures. With the development of FinFET quantum dots and increased operating temperatures of spin qubits ($\sim$\SI{1}{\kelvin})~\cite{yang2020operation,petit2020universal}, it may be possible to fully integrate the quantum processor with the classical controller on-chip or by flip-chip technology, thus lifting a major roadblock in scaling.\\
\linebreak
\textbf{Data availability}
Data supporting this work will be uploaded to online repository.\\
\linebreak
\textbf{Acknowledgements}
This research was funded by the Intel Corporation. We acknowledge useful discussions with the members in the Spin Qubit team, the Cryo-CMOS team and Intel Corporation, and technical assistance by O. Benningshof, M. Sarsby, R. Schouten and R. Vermeulen.\\
\linebreak
\textbf{Author contributions}
X.X., B.P. and J.P.G.D performed the experiment. N.S. fabricated the quantum device. A.S. and G.S. grew the Si/SiGe heterostructure. A.C. contributed to the preparation of the experiment. X.X. and B.P. analysed the data. F.S., M.B., S.P., E.C. and L.M.K.V. conceived and supervised the project. X.X., B.P. and L.M.K.V. wrote the manuscript with input from all authors.\\
\linebreak
\textbf{Competing interests}
The authors declare no competing interests.
\bibliography{references}
% \pagebreak
\clearpage
\section*{Methods}

\setlength{\parindent}{0em}
\textbf{Programming the cryo-controller.}
The setup (Extended Data Fig.\ref{fig:setup_2}) contains a field-programmable gate array (FPGA) that configures the cryo-controller (e.g. FTW), programs the various memories inside the cryo-controller (e.g. envelope memories, instruction tables and instruction lists), and controls the start of the execution of the instruction list. The FPGA is connected to the host PC, which sends the data that needs to be uploaded to the cryo-controller over the SPI. The instruction list integrated in the cryo-controller does not support classical instructions that allow for e.g. branching or wait statements, as required for performing certain qubit experiments and for synchronization with other equipment. Therefore,  switching between different instruction lists and synchronization with the rest of the equipment, is controlled by two trigger lines from the AWG to the FPGA. The application of the \emph{execute} trigger starts the execution of the instruction list that is programmed in the cryo-controller, for performing repeated measurements. The application of the \emph{sweep} trigger loads the next instruction list from the FPGA SRAM into the cryo-controller's instruction list.\\
\textbf{Power budget.} The high power consumption of the digital circuitry of the cryo-controller is caused due to the lack of clock gating in registers (memory), thus causing them to continuously operate instead of just during the read/write cycle. This could easily be reduced by further optimizations (e.g. by replacing more registers with SRAM memory and by adding clock gating), that were not yet included in the first generation cryo-controller. Based on the simulation with clock-gating, the power consumption of the digital circuitry should be lower than \SI{40}{\milli\watt} instead of \SI{330}{\milli\watt} in the current design. Moreover, this chip was designed to address both transmons and spin qubits and hence an ultra-wide output frequency range was supported i.e. 2 to \SI{20}{\giga\hertz}. Once the qubit frequency is fixed within a few GHz range, the power consumption of the analog circuitry can be significantly reduced to limit the power consumption to $\sim$ \SI{20}{\milli\watt} instead of \SI{54}{\milli\watt}.\\
\textbf{Si/SiGe heterostructure.}
The quantum processor is made by gate-confined quantum dots in a $^{28}$Si/SiGe heterostructure, which is grown using a reduced pressure chemical vapor deposition reactor (ASM Epsilon 2000). First, a Si$_{1-x}$Ge$_{x}$ buffer layer (with $x$ linearly increasing from 0 to 0.3) is grown on top of a p-type natural Si wafer, followed by a \SI{300}{\nano\meter} strain-relaxed Si$_{0.7}$Ge$_{0.3}$ layer. Then a \SI{10}{\nano\meter} isotopically purified tensile-strained $^{28}$Si (with \SI{800}{ppm} residual $^{29}$Si concentration) quantum well is grown, followed by a \SI{30}{\nano\meter} strain-relaxed Si$_{0.7}$Ge$_{0.3}$ barrier layer. Finally a \SI{1}{\nano\meter} sacrificial Si cap is grown on top.\\
\textbf{Quantum dot device fabrication} On top of the heterostructure, a \SI{7}{\nano\meter} AlO$_x$ layer is deposited using atomic-layer deposition (ALD), followed by a \SI{20}{\nano\meter} Al metal film, which is patterned using electron beam lithography in order to define a first gate layer, which shapes the potential landscape. Next another \SI{7}{\nano\meter} AlO$_x$ layer is deposited, followed by a \SI{70}{\nano\meter} Al layer which uniformly covers the quantum dot area. Finally, a \SI{200}{\nano\meter} Co film is deposited and patterned into a  micro-magnet (see Extended Data~\cref{fig:micromagnet}).\\
\textbf{Qubit readout.} 
The readout scheme is described in Extended Data Fig.\ref{fig:sequence}. After each operation sequence, $Q_2$ is measured by spin-selective tunneling to the electron reservoir, where a spin-up ($\ket{1}$) electron can tunnel out and a spin-down ($\ket{0}$) electron is blockaded from tunneling out. Such a spin-to-charge conversion changes the charge occupancy in the quantum dot conditional on the spin state. This in turn changes the current signal in an adjacent capacitively coupled single-electron-transistor (SET). Single-shot readout of the qubit state can be done by thresholding the current signal through the SET~\cite{xue2020repetitive}. The post-measurement state in this readout protocol is the $\ket{0}$ state, serving as reinitialization. $Q_1$ is tuned to be only weakly coupled to the SET, which serves as the electron reservoir for $Q_1$. This is to minimize the back-action from the SET, but also makes it less efficient to readout $Q_1$ by spin-selective tunneling to the SET. Therefore, with $Q_2$ reinitialized, a $CROT$ gate is applied to map the state of $Q_1$ onto $Q_2$. Then $Q_1$ is readout by measuring $Q_2$ again~\cite{xue2020repetitive}. The readout fidelity of $Q_2$ is mainly limited by the thermal broadening of the electron reservoir, and the readout fidelity of $Q_1$ is limited by both the error in the $CROT$ gate and in the readout of $Q_2$. Thus the readout visibility of $Q_1$ is lower than $Q_2$.\\
\textbf{Readout error removal.}
In the AllXY experiments and in the implementation of the Deutsch-Jozsa algorithm, the readout probabilities of $Q_2$ are normalized with calibrated readout fidelities $(F_{\ket{0}}, F_{\ket{1}})$. 
After preparing $Q_2$ in $\ket{0}$, $F_{\ket{0}}$ can be calibrated directly through the measured spin-down probability, and $F_{\ket{1}}$ is calibrated through the measured spin-up probability after a spin-flip operation (the spin-flip fidelity is above $99\%$). Based on the measured state probabilities in the AllXY and Deutsch-Jozsa experiments, $\bm{P^M} = (P_{\ket{0}}^M, P_{\ket{1}}^M)^T$, the actual state probabilities $(P_{\ket{0}}, P_{\ket{1}})$ can be reconstructed by $\bm{P} = \bm{F}^{-1}\bm{P^M}$, where 
\begin{equation}
\bm{F}=
\begin{pmatrix*}
	F_{\ket{0}} & 1 - F_{\ket{1}}\\
	1 - F_{\ket{0}} & F_{\ket{1}}
\end{pmatrix*}.
\end{equation}
\textbf{Error sources.} 
In the simultaneous Rabi oscillation experiment (Fig.\ref{fig:single-qubit}), we attribute the visible decays in both curves to the residual exchange coupling between the two qubits. Simultaneous Rabi oscillations recorded (in this case using the RT setup) over larger numbers of oscillations show beating patterns. These patterns are well reproduced by numerical models of the spin evolution in the presence of a finite residual exchange coupling. Such a beating effect looks like a decay in the beginning. It is absent in the individually driven Rabi oscillation (Extended Data Fig.\ref{fig:individual_rabi}). In the two-qubit experiments shown in Fig.\ref{fig:two-qubit}, the decay in the controlled-rotation Rabi oscillations and the finite visibilities in the Deutsch-Jozsa algorithm are largely attributed to charge noise. With the exchange coupling turned on, as needed for two-qubit gates, the energy levels are much more sensitive to charge noise.\\
\textbf{Quantum state tomography.}
In the QST experiment, the qubit state is measured by projecting it onto the $(-\hat{z},+\hat{x},-\hat{y},+\hat{z})$ axes. The projection on the $-\hat{z}$ axis is measured by direct readout of the spin state, while the projections on other axes are measured by applying a $X$, $Y$, or $X^2$ gate, which are calibrated by the AllXY experiment, before the readout. The trajectory of the qubit state in the course of a $X^2$ gate can be reconstructed by performing QST at incremental burst times of a rectangular microwave signal~(Fig.\ref{fig:single-qubit}.c.), with each measurement repeated 1,000 times. To visualize the qubit state in the Bloch sphere, we remove the readout error from the data. Since error removal can lead to unphysical states such as data points outside the Bloch sphere, a maximum likelihood estimation is implemented to find the closest physical state of the qubit~\cite{altepeter2005photonic}.\\
\renewcommand{\figurename}{Extended Data Fig.}
\setcounter{figure}{0}

\begin{figure*}[t]  
    \center{\includegraphics[width=1\linewidth]{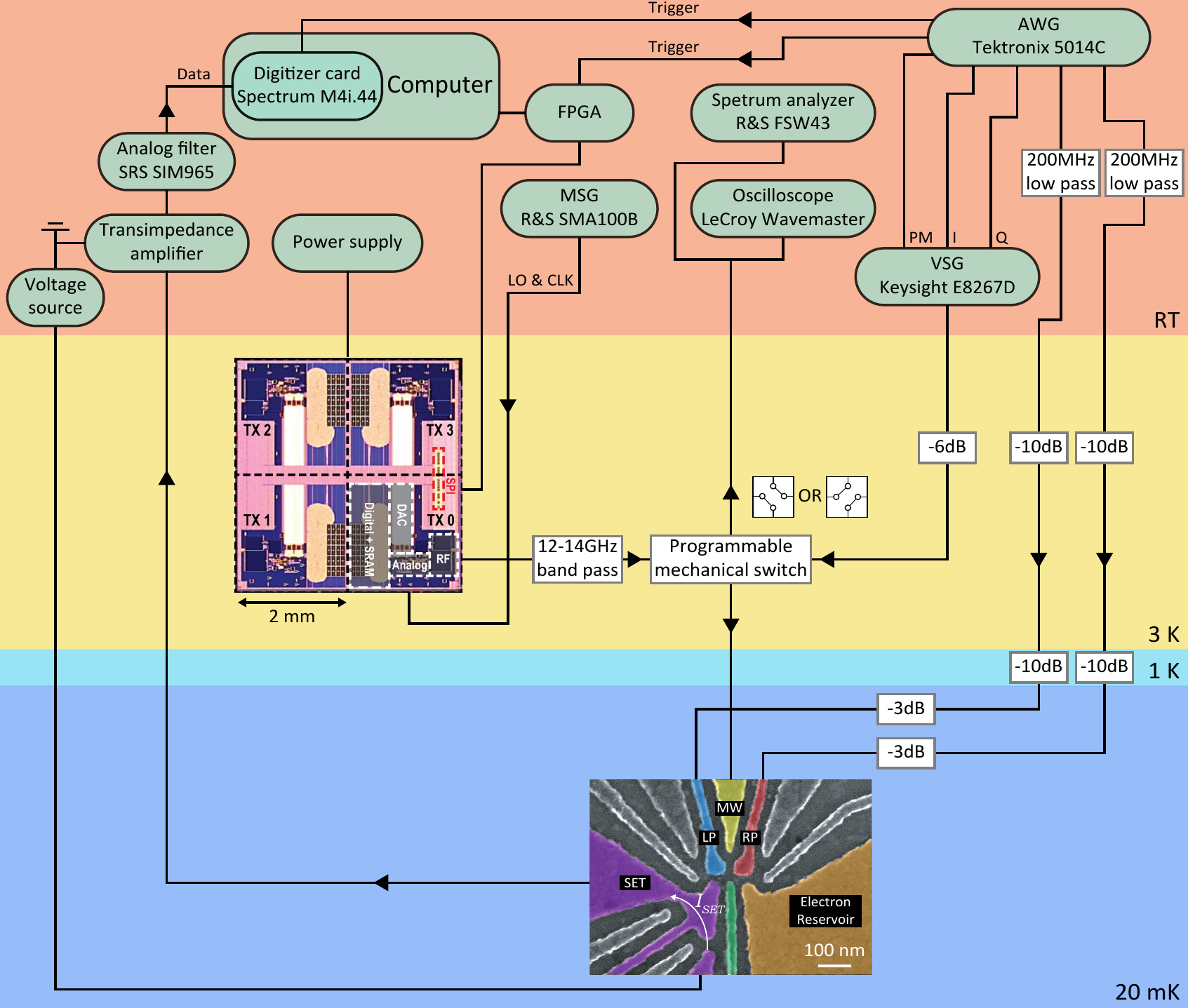}}
    \caption{\textbf{Experiment setup.} The quantum dot device is wire-bonded onto a printed circuit board (PCB) which is placed at the mixing chamber ($\sim$\SI{20}{\milli\kelvin}) of a dilution refrigerator (Bluefors XLD). Voltage pulses onto gates RP and LP are generated by the AWG at room temperature (RT), and go through a low-pass filter (Minicircuits) and attenuators before reaching the device. These pulses are used to control the electrochemical potentials of the quantum dots and load/unload electrons from/to the electron reservoir (See Extended Fig.~\ref{fig:sequence}). A programmable mechanical switch at \SI{3}{\kelvin} is used to connect gate MW either to a vector signal generator (VSG) at RT or to the cryo-controller at \SI{3}{\kelvin} (represented as two boxes next to the switch) through a 12-\SI{14}{\giga\hertz} band-pass filter to filter out wide-band noise. The mechanical switch can also be configured to send the output signals from the cryo-controller to the oscilloscope and the spectrum analyzer at RT for electrical characterization in time and frequency domain. The cryo-controller is programmed via an FPGA to generate the microwave bursts using an external local oscillator (LO) signal and a clock (CLK) signal from a microwave signal generator (MSG) at RT. The single electron transistor (SET) next to the quantum dots is voltage biased and the current signal ($I_{SET}$) through it is converted to a voltage signal through a transimpedance amplifier and digitized by a digitizer card after a \SI{10}{\kilo\hertz} analog low-pass filter. $I_{SET}$ is sensitive to the charge occupation of the quantum dots, allowing binary single-shot readout of the qubit states via spin-to-charge conversion (\cref{fig:sequence}).}
    \label{fig:setup_2}
\end{figure*}  

\begin{figure*}[t] 
    \center{\includegraphics[width=0.9\linewidth]{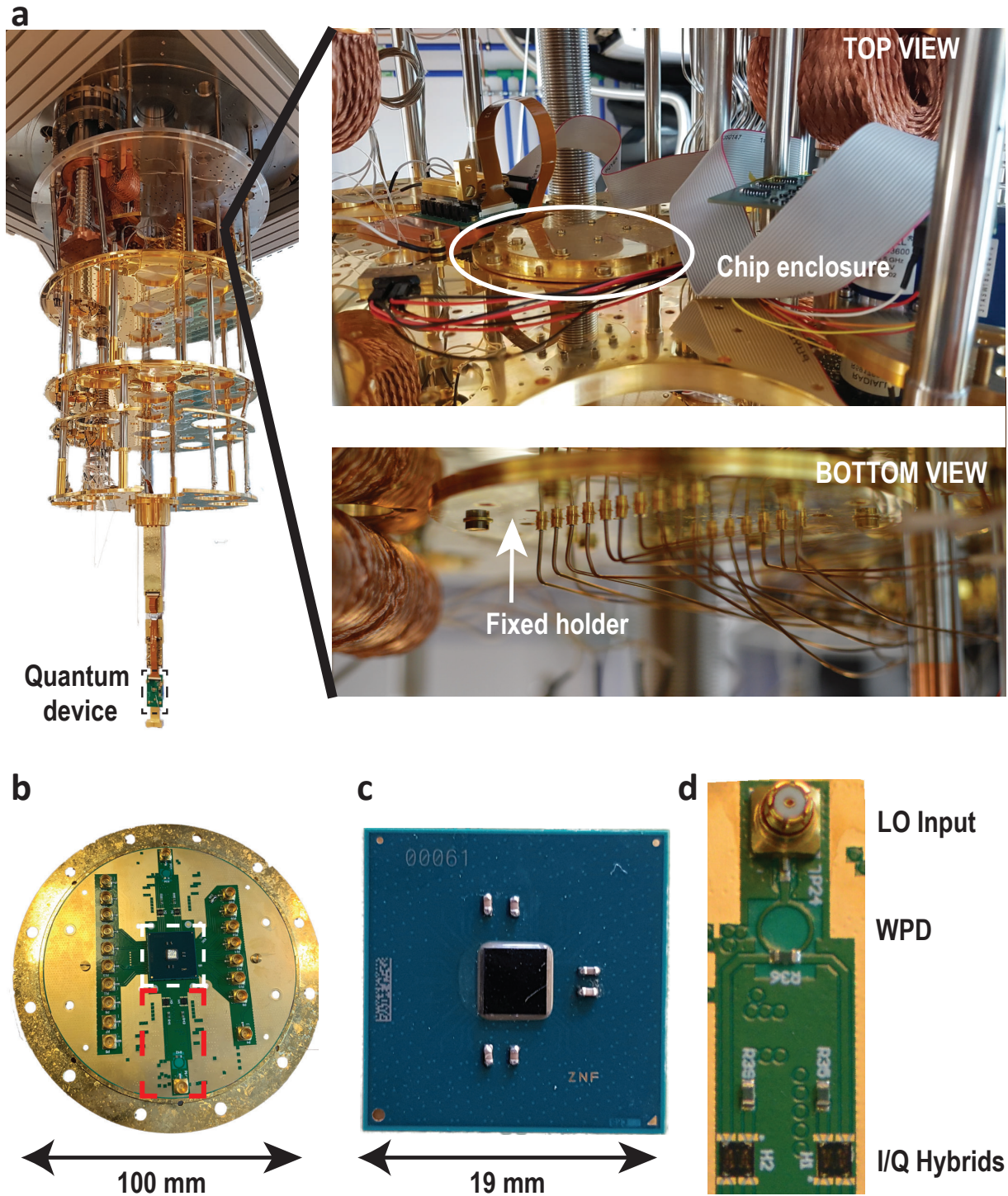}}
    \caption{\textbf{Dilution refrigerator setup} \textbf{a.} Location of the cryo-controller and the quantum device inside the dilution refrigerator (left). Top view and bottom view of the \SI{3}{\kelvin} plate showing the mounted chip enclosure and the fixed holder for the enclosure respectively (right). \textbf{b.} Top view of the gold-plated annealed copper enclosure (without the lid) which is used to mount and thermalize the cryo-controller. \textbf{c.} Ball-grid-array(BGA)-324 package hosting the cryo-controller chip with on-package decoupling capacitors (highlighted as white box in \textbf{b}). \textbf{d.} The Wilkinson power divider (WPD) splits the input LO power into two equal paths with half power in each implemented on a PCB. Discrete I/Q hybrids create the in-phase and quadrature phase component of the input LO are wire-bonded on the PCB for LO distribution between the different transmitters inside the cryo-controller (highlighted as red box in \textbf{b}).}
    \label{fig:fridge_setup}
\end{figure*}  

\begin{figure*}[t] 
    \center{\includegraphics[width=1\linewidth]{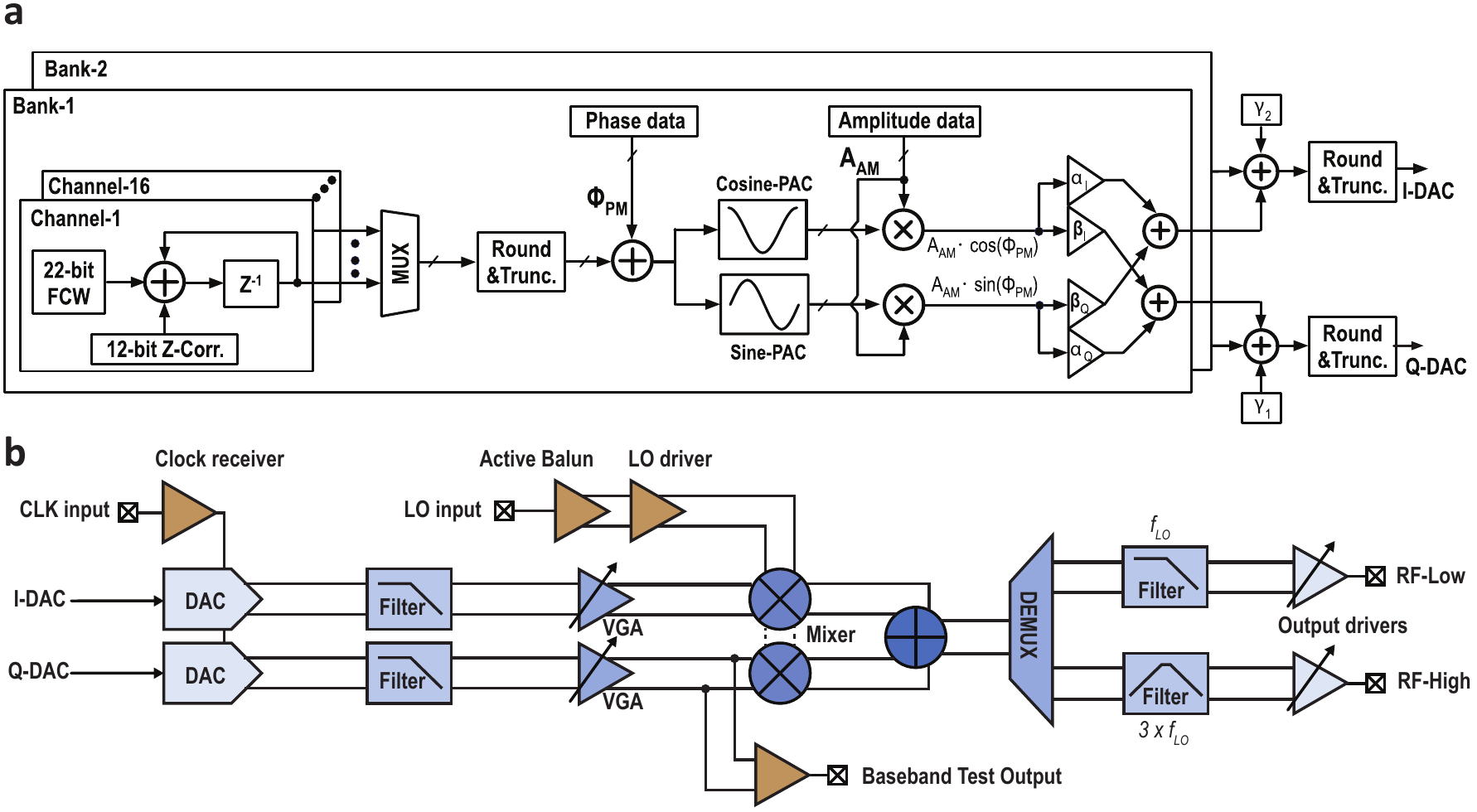}}
    \caption{\textbf{Detailed cryo-controller schematic} \textbf{a.}\,Detailed representation of the digital circuitry. \textbf{b.}\,Detailed system-level schematic of the analog circuity inside the controller.}
    \label{fig:schematic}
\end{figure*}  

\begin{figure*}[t] 
    \center{\includegraphics[width=1\linewidth]{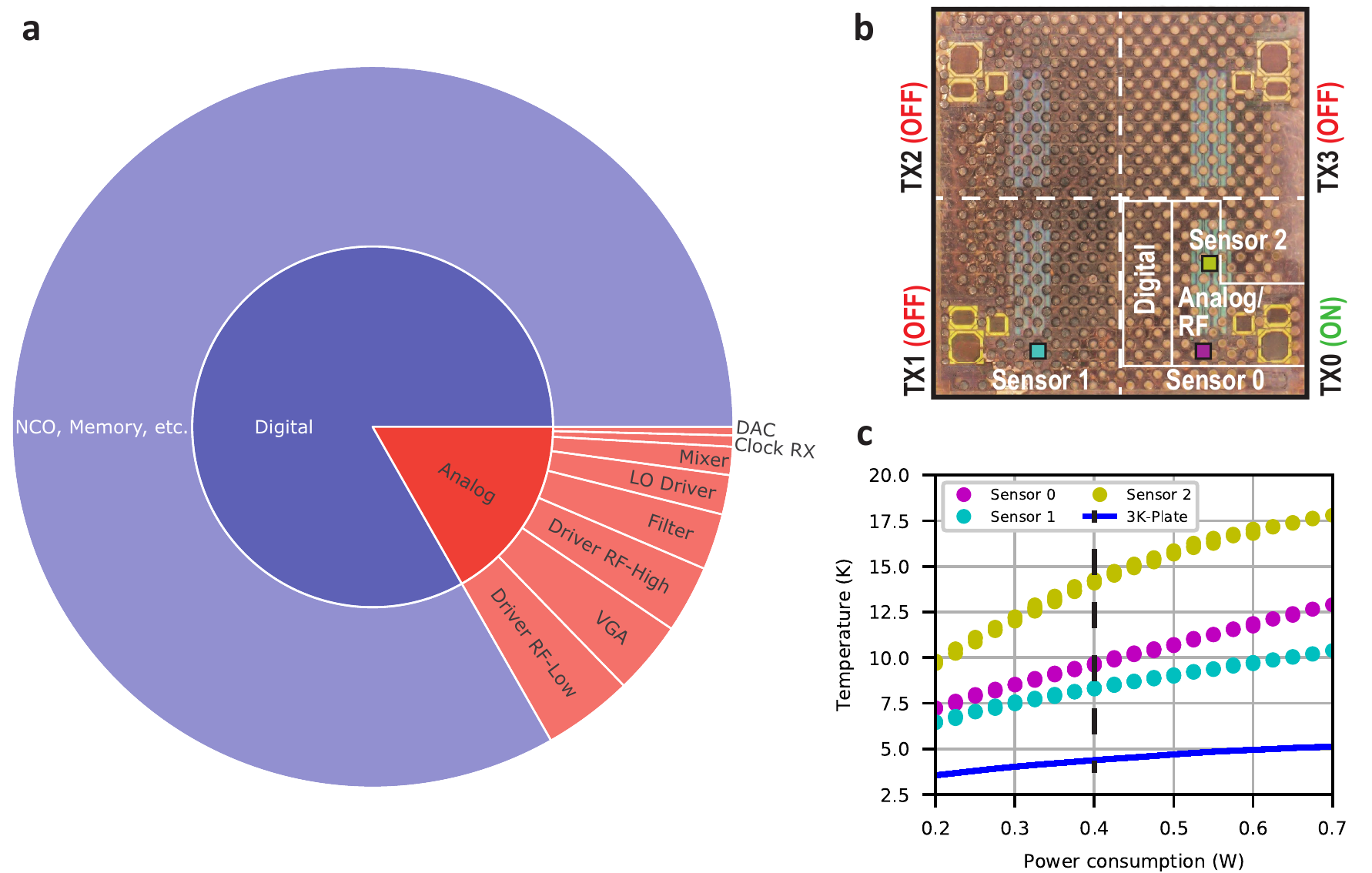}}
    \caption{\textbf{Power consumption and self-heating of the cryo-controller.} \textbf{a.} Power consumption pie chart showing the contribution of the digital and analog circuits in the cryo-controller. The power-consumption breakdown of individual circuit blocks are shown for the analog circuits. \textbf{b.} Chip micrograph showing the on-chip bumps used as I/Os. The locations of on-chip temperature sensing diodes and the analog and digital circuitry (in TX0) are highlighted. \textbf{c.} The measured on-chip and \SI{3}{\kelvin} plate temperature using different sensors versus the power consumption of TX0, as reported in \cite{patra202019}. The power consumption is varied by changing the clock frequency of the chip. The nominal operating point for the work presented here and corresponding temperatures are highlighted with a dashed vertical line. All the other transmitters (TX1, TX2, TX3) are switched OFF in this measurement.}
    \label{fig:power}
\end{figure*}

\begin{figure*}[t] 
    \center{\includegraphics[width=1\linewidth]{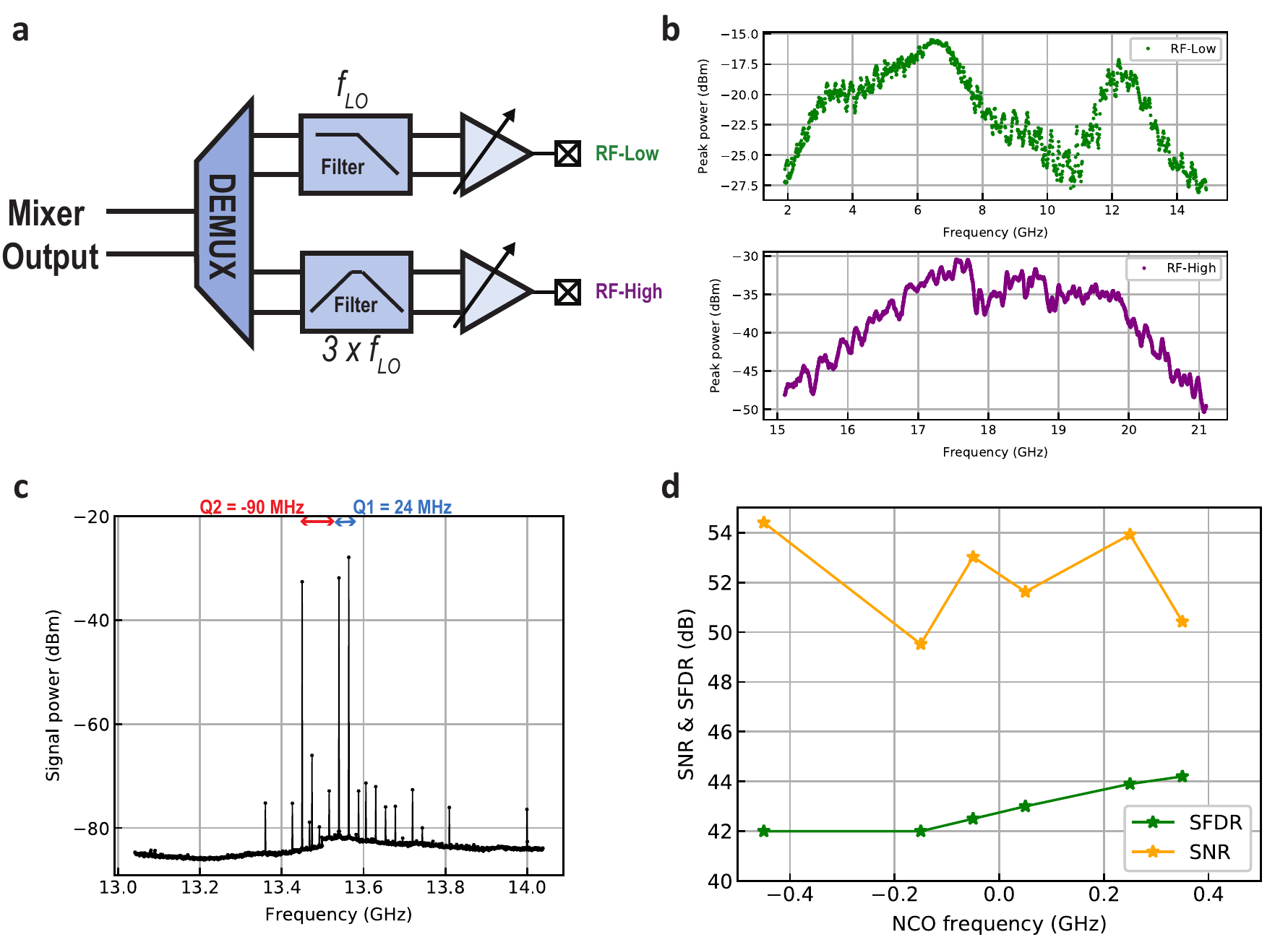}}
    \caption{\textbf{Detailed electrical characterization of the cryo-controller.} \textbf{a.} Schematics of the output driver (complete version in Extended Data~\cref{fig:schematic}) showing the two different RF outputs which use the same external LO to generate two different frequencies i.e. a \SI{1}{\giga\hertz} band around $f_{LO}$ or a \SI{1}{\giga\hertz} band around $3\times f_{LO}$ by selecting the RF-Low or RF-High path respectively. \textbf{b.} Peak output power versus frequency generated using the RF-Low and RF-High path respectively, as reported in \cite{patra202019}. \textbf{c.} Two-tone output spectrum of the cryo-controller used in the simultaneous Rabi oscillation experiment. \textbf{d.}\,SNR and SFDR of the cryo-controller at various NCO frequencies around \SI{13.54}{\giga\hertz}.}
    \label{fig:rf_output}
\end{figure*}   

\begin{figure*}[t]  
    \center{\includegraphics[width=1\linewidth]{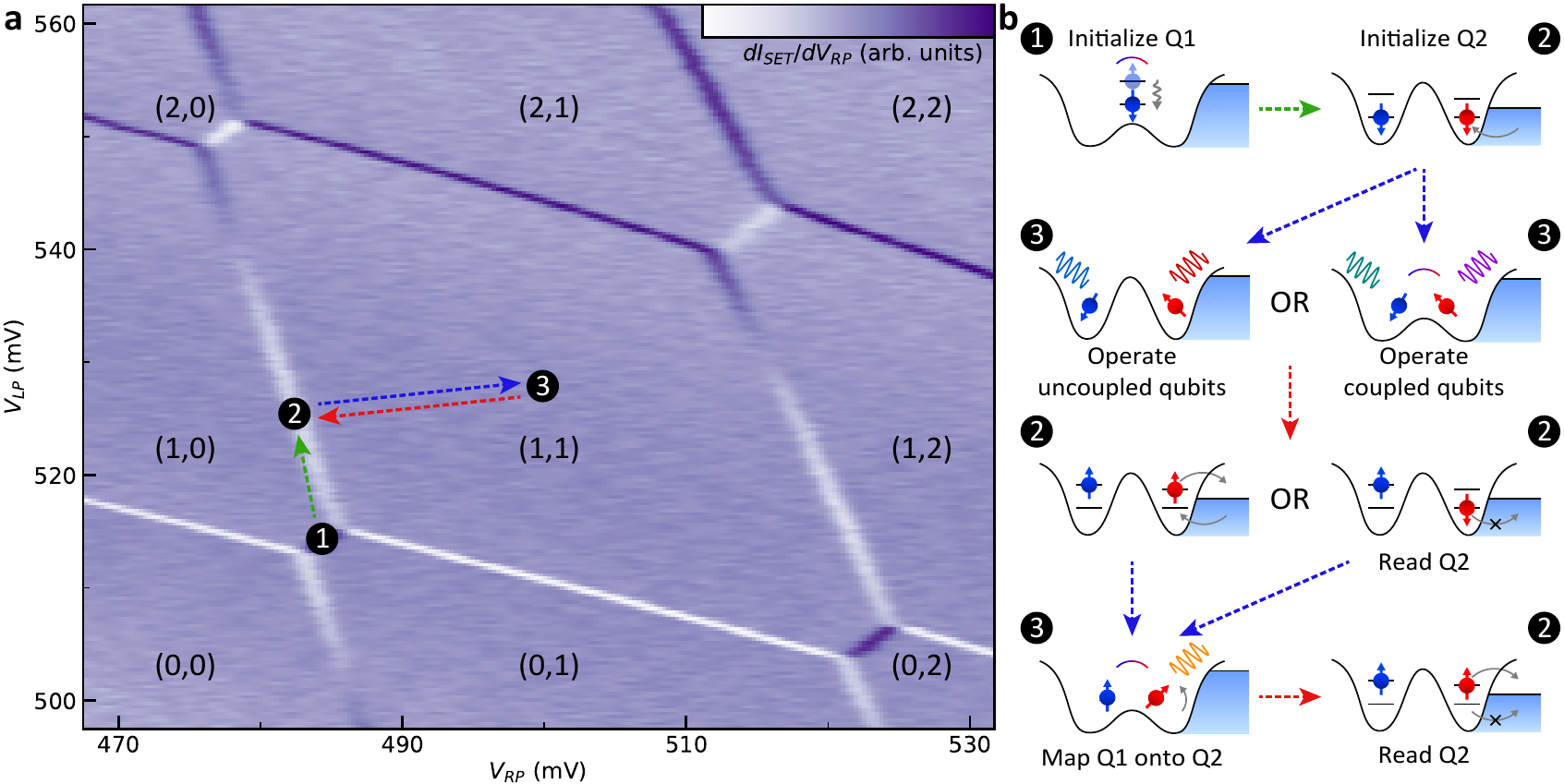}}
    \caption{\textbf{Pulsing scheme in qubit experiments.} \textbf{a.} Charge stability diagram of the DQD system, showing the differential current signal ($dI_{SET}/dV_{RP}$) and charge occupation ((M, N), indicating M electrons in the dot below LP and N electrons in the dot below RP) as a function of the voltages applied to gate LP ($V_{LP}$) and gate RP ($V_{RP}$). The three main stages of a typical pulse sequence are marked by the numbered circles. The gate voltages of stage 3 vary between different experiments: in the experiments with exchange coupling turned on, due to the cross-capacitance between the barrier (gate T) and the plungers (gates RP and LP), the LP and RP voltages are different from the experiments without exchange coupling by $\sim$\SI{15}{\milli\volt}. \textbf{b.} Schematic representations of the DQD system during the experiment cycle. $Q_1$ is first initialized to its ground state (spin-down) via fast relaxation by pulsing to the charge transition line between (1,0) and (0,1) (stage 1), which is a spin-relaxation hotspot~\cite{srinivasa2013simultaneous}. Then $Q_2$ is initialized by pulsing it to the transition line between (1,0) and (1,1) (stage 2), where the Fermi energy of the electron reservoir is placed in between the two spin states of $Q_2$. It allows a spin-down electron to tunnel into the dot but forbids spin-up electrons from tunneling in, a mechanism called spin-selective tunneling. During the qubit operations, the system is pulsed to the middle of the (1,1) region (stage 3) so both electrons are well-confined inside the DQD. The barrier (gate T) voltage is used to turn off the exchange coupling between the two spins in the operation of uncoupled qubits (all measurements in \cref{fig:single-qubit}) and to turn on the coupling for two-qubit logic operations (all measurements in \cref{fig:two-qubit}). After the operations, $Q_2$ state is read out via spin-selective tunneling and reinitialized into the spin-down state (stage 2). The state of $Q_1$ is read out by mapping its state onto $Q_2$ via a two-qubit $CROT$ gate (stage 3), followed by readout of $Q_2$ again (stage 2).
    }
    \label{fig:sequence}
\end{figure*}  
\pagebreak
\begin{figure*}[t] 
    \center{\includegraphics[width=1\linewidth]{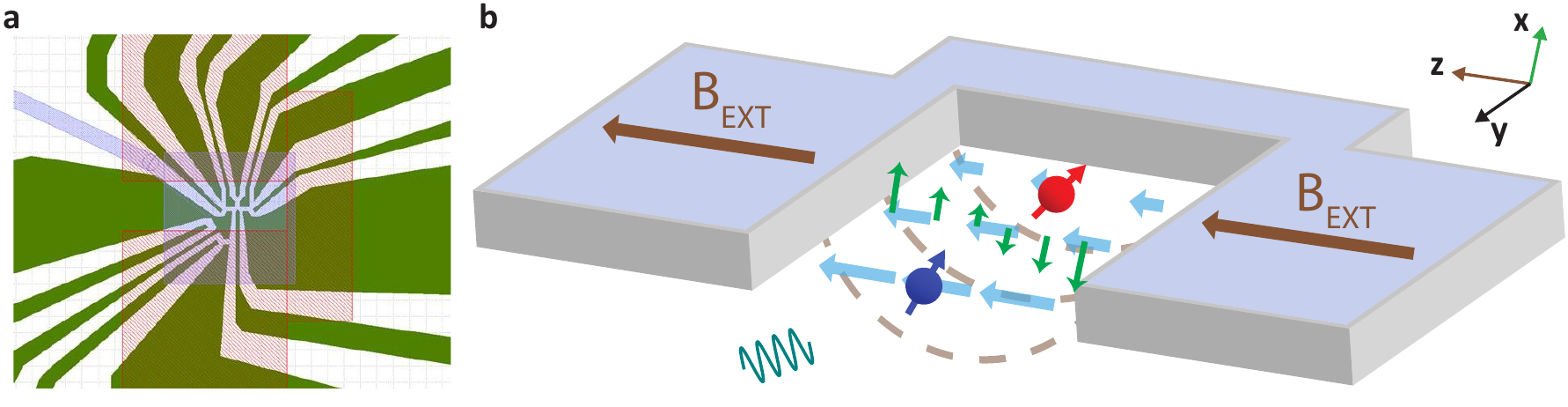}}
    \caption{\textbf{Magnetic field gradient.}  \textbf{a.} Schematic showing the first and second Al gate layers in green and purple, respectively. A Cobalt micro-magnet is located on top of the metallic gates (light red shaded area). \textbf{b.} The micromagnet is magnetized by sweeping the external magnetic field (in $\hat{z}$ direction) from 0 to \SI{3}{\tesla} and back to \SI{380}{\milli\tesla}. The magnetized micro-magnet provides an additional magnetic field (brown dashed lines) which has a longitudinal ($\hat{z}$) component with a field gradient along the double quantum dots. This longitudinal magnetic field gradient (light blue arrows) makes the Zeeman splittings (resonance frequencies) of the two qubits different by $\sim$\SI{110}{\mega\hertz}. Additionally, the micro-magnet also induces a transverse ($\hat{x}$) magnetic field gradient (green arrows). When a microwave pulse is sent to the device through gate MW, the wave functions of the electrons are oscillating in the $\hat{z}$ direction. If the microwave frequency is on resonance with the qubit frequency, the electron is subject to an oscillating magnetic field along the $\hat{x}$ direction, which induces electric-dipole spin resonance (EDSR)~\cite{pioro2008electrically}.
    }
    \label{fig:micromagnet}
\end{figure*}  
\pagebreak

\begin{figure*}[t]  
    \center{\includegraphics[width=1\linewidth]{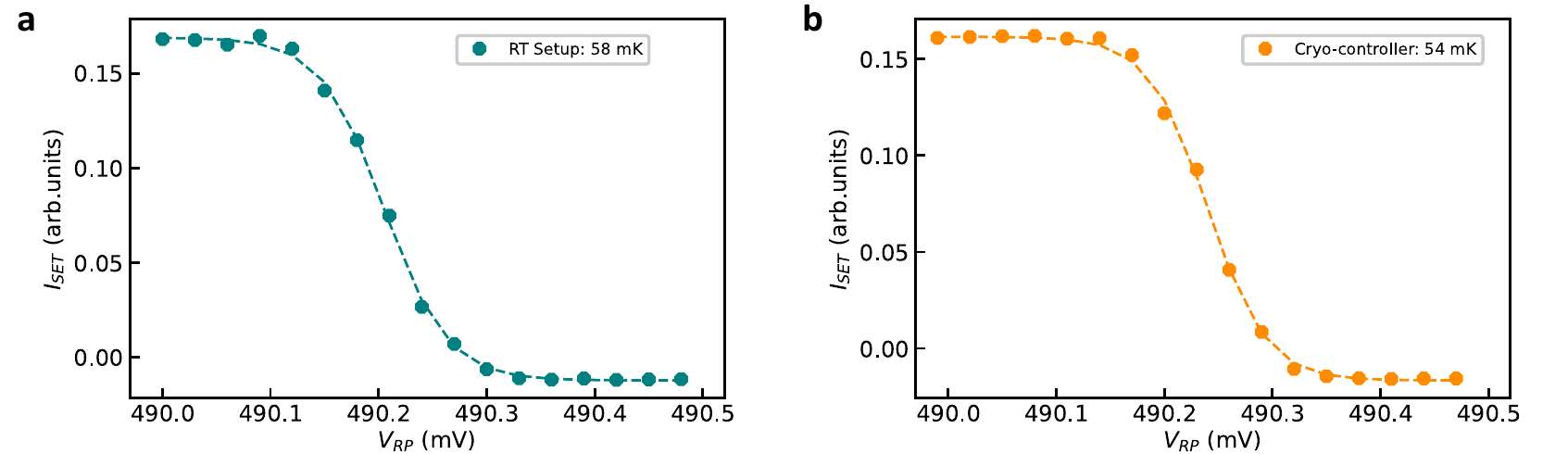}}
    \caption{\textbf{Electron temperature measured at different configurations.} SET current signal ($I_{SET}$) as a function of RP voltage ($V_{RP}$) measured at the charge transition between (1,0) and (1,1) when the quantum device is connected to the VSG (\textbf{a}) and to the cryo-controller (\textbf{b}) (at zero magnetic field). The electron temperatures are extracted by fitting the curves with the Fermi-Dirac distribution, with a lever arm of \SI{0.172}{\milli{eV/V}}. The measurements indicate that the output noise of the cryo-controller does not affect the electron temperature.
    }
    \label{fig:electron_temperature}
\end{figure*}    
\pagebreak    
\begin{figure*}[t] 
    \center{\includegraphics[width=1\linewidth]{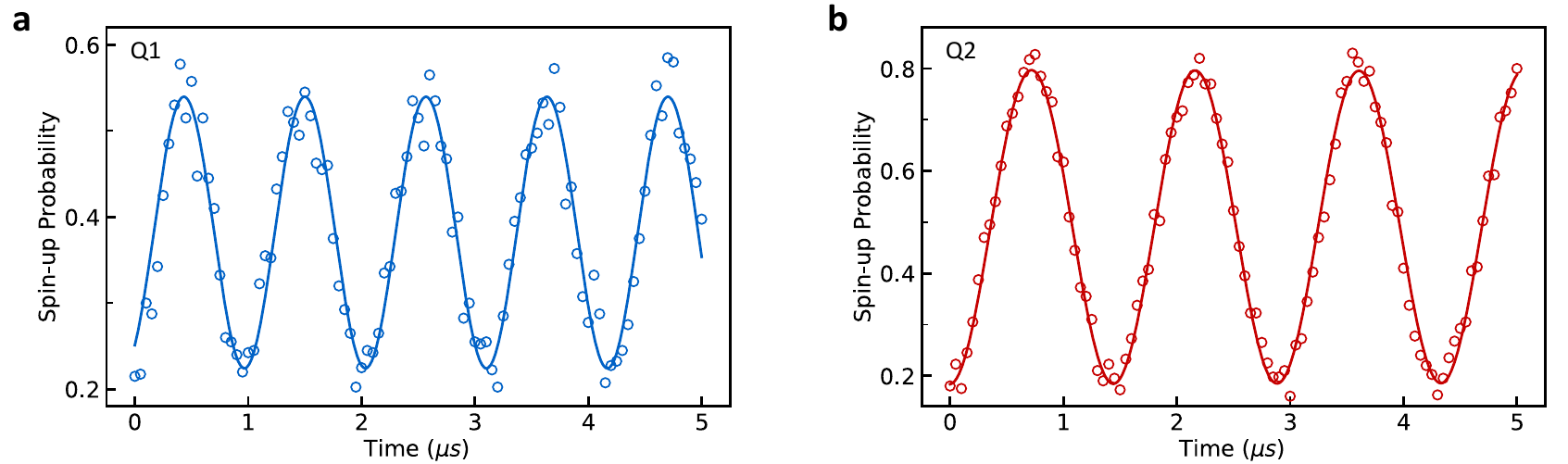}}   
    \caption{\textbf{Rabi oscillations of qubits individually driven by the cryo-controller.} The output frequency of two NCOs are set to the frequencies of $Q_1$ and $Q_2$ respectively, but only one NCO is active each time. Using the same method as described in the main text, Rabi oscillations of $Q_1$ (\textbf{a}) and $Q_2$ (\textbf{b}) are measured individually. Compared to the simultaneous Rabi oscillations shown in \cref{fig:single-qubit}.b, the decay is much slower in the individual driving experiments.
    }
    \label{fig:individual_rabi}
\end{figure*}

\end{document}